\documentclass{eceasst}

\usepackage{subcaption}
\usepackage{pdflscape}  
\usepackage{longtable}  
\usepackage{hyperref}

\title{\texorpdfstring{Insights into User Interface Innovations from a \newline Design Thinking Workshop at deRSE25}{Insights into User Interface Innovations from a Design Thinking Workshop at deRSE25}}
\short{Design Thinking Workshop at deRSE25}
\author{\texorpdfstring{Maximilian Frank\autref{1}$^*$ and Simon Lund\autref{2}$^*$}{Maximilian Frank and Simon Lund}}
\institute{
\autlabel{1}\email{maximilian.frank@psy.lmu.de} (ORCID: \href{https://orcid.org/0000-0002-8140-3519}{0000-0002-8140-3519})\\ 
\autlabel{2}\email{simon.lund@lmu.de} (ORCID: \href{https://orcid.org/0009-0006-5907-1577}{0009-0006-5907-1577})\\
\vspace{0.2cm}
\href{https://ror.org/05591te55}{Ludwig-Maximilians-Universität München} \\
Munich, Germany\\
\vspace{0.4cm}
\textit{*both authors contributed equally}\\
\vspace{0.8cm}
Submitted to \href{https://eceasst.org/index.php/eceasst/about}{\textit{ECEASST}}  (Electronic Communications of the EASST) on 18th May 2025}

\abstract{
Large Language Models have become widely adopted tools due to their versatile capabilities, yet their user interfaces remain limited, often following rigid, linear interaction paradigms. In this paper, we present insights from a design thinking workshop held at the deRSE25 conference aiming at collaboratively developing innovative user interface concepts for LLMs. During the workshop, participants identified common use cases, evaluated the strengths and shortcomings of current LLM interfaces, and created visualizations of new interaction concepts emphasizing flexible context management, dynamic conversation branching, and enhanced mechanisms for user control. We describe how these participant-generated ideas advanced our own whiteboard-based UI approach. The ongoing development of this interface is guided by the human-centered design process - an iterative, user-focused methodology that emphasizes continuous refinement through user feedback. Broader implications for future LLM interface development are discussed, advocating for increased attention to UI innovation grounded in user-centered design principles.}

\begin{document}
\maketitle

\section{Introduction}
The era of Large Language Models (LLMs) is rooted in the publication of \textit{Attention Is All You Need} \cite{attention-is-all-you-need} in 2017, which introduced the transformer architecture for machine translation tasks.
While the underlying components were not entirely new -- they build upon years of basic research in fields in deep learning -- the paper synthesized them into an effective design compared to existing architectures (e.g., recurrent neural networks). 
The broader impact of transformers and widespread adoption, however, came only after OpenAI released its GPT-3 model along with a web interface. 
This interface allowed users to interact with LLMs and brought the architecture into mainstream use \cite{zhao2025surveylargelanguagemodels, naveed2024comprehensiveoverviewlargelanguage}.

Today LLMs are used across diverse applications due to their versatile range of capabilities: world knowledge, multilingual support, text comprehension, instruction following, in-context learning, reasoning, user interaction, self-improvement, and tool utilization \cite{minaee2025largelanguagemodelssurvey}.
Applications include question-answering assistants, multi-step reasoning agents, creative content generation, programming support, data analysis, research tools, language learning, and integration into games \cite{kaddour2023challengesapplicationslargelanguage, ma2023demonstrationinsightpilotllmempoweredautomated,Hadi_2023,KASNECI2023102274,10680313}.

While the capabilities of LLMs are rapidly evolving, the evolution of UI has not kept pace.
Most LLM interfaces still follow a rigid conversation structure constrained by sequential integrity requirements -- preventing contextual restructuring and masking parts of the conversation.
This may increase the likelihood of model hallucinations, e.g., due to unalterable inconsistencies or focus shifts across multiple topics across extended interactions.
For instance, in development contexts, when dialogue involves peripheral discussions -- explaining code, identifying bugs, or managing feature requests -- models struggle to discern and prioritize essential information.
Quality may further degrade when conversations exceed context length limitations, resulting in information loss through truncation or compression.
As a consequence, users may restart conversations and reconstruct context, potentially disrupting workflow continuity and affecting trust in model outputs.

In this paper, we present the results of a design thinking workshop at deRSE25 on the topic of user interface development for LLMs. 
After a brief introduction into the operation principles of LLMs, we highlight the strengths of a more user-driven development for LLM interfaces and discuss -- in our perception currently unutilized -- potentials of interface development.
\section{Background and Motivation}

\subsection{Functioning of LLMs}
LLMs are stochastic models that predict the next token based on preceding input \cite{bender2021dangers}. 
A token is a basic unit of text that may correspond to part of a word, a complete word, or multiple words depending on the tokenization algorithm \cite{kudo2018sentencepiece}.
For example, the word "unbelievable" might be split into tokens like "un", "believe", and "able".

\begin{figure}
    \centering
    \includegraphics[width=0.3\linewidth]{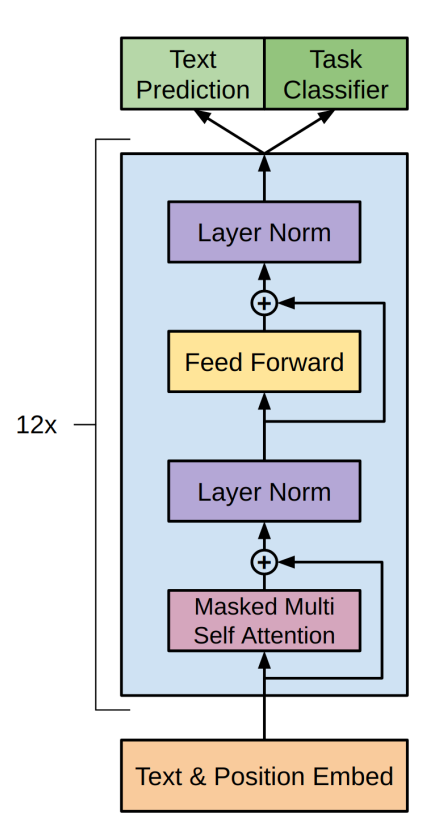}
    \caption{Decoder-only architecture showing how LLMs process input tokens through multiple transformer blocks to generate output tokens sequentially. The self-attention mechanism allows each token to attend to all previous tokens in the context window. \cite{radford2018improving}}
    \label{fig:decoder-architecture}
\end{figure}

An LLM does not maintain internal state between interactions. 
Instead, it operates on a fixed-length context window containing the current prompt, conversation history, and system prompt (instructions about its role and answer behavior).
Context length limits stem from hardware constraints (computational resources, VRAM requirements) and training limitations (maximum token counts models process). 
When processing input, the model applies transformer decoder blocks with self-attention mechanisms (Figure \ref{fig:decoder-architecture}) to generate output token-by-token until producing a stop token. 
The generated response is then appended to the context window for subsequent interactions.

The model's capability derives from parameters tuned during pre-training on large text corpora \cite{pile}, encoding linguistic patterns, factual knowledge, and reasoning capabilities that enable contextually meaningful responses.
However, this is no guarantee for factual accuracy or logical consistency, as models may generate plausible-sounding but incorrect or inconsistent information, commonly referred to as hallucinations\footnote{There is ongoing debate regarding this terminology. "Hallucination" medically refers to "the perception of an entity or event that is absent in reality" (\cite{macpherson2013hallucination} as cited in \cite{huang2025survey}). Some scholars suggest alternative terms like "bullshitting" for inaccurate AI outputs \cite{hicks2024chatgpt}.} \cite{huang2025survey}.

\subsection{Limitations of Current LLM User Interfaces}
Recent research has extensively documented various usability challenges and limitations inherent in current Large Language Models (LLMs). 
Prominent among these issues are the persistent occurrences of hallucinations, misinformation, and biased outputs, where LLMs confidently generate plausible yet incorrect information \cite{bender2021dangers, shen2024thermometer}.

Such inaccuracies are exacerbated by inadequate calibration of trust and user confidence, leading users to either excessively trust erroneous responses or to mistrust correct ones \cite{Vafa2024HumanGeneralization, Metzger2024CalibratedTrust}.

Further compounding these challenges are mismatches between users' mental models and the operational realities of LLMs, often causing confusion, misuse, or unmet expectations \cite{Grimes2021MentalModels, Wang2025MentalModels}. 
Additionally, accessibility and inclusivity remain significant concerns, with LLMs frequently perpetuating biases or failing to accommodate diverse user needs, further hindering their widespread, equitable adoption \cite{Gadiraju2023DisabilityLLMs, Martinez2024EasyRead}.

While a considerable body of work focuses on enhancing the technical and functional reliability of LLMs through methods such as calibration models and external knowledge integration \cite{shen2024thermometer, Pak2024PoliteAI}, limited research has specifically addressed improvements to the user interface (UI) through which end-users interact with these systems. 

The dominant paradigm in LLM interaction is the linear chat interface, where conversations progress sequentially without structural modification. 
Such interfaces limit users to a single conversation thread, with each message building linearly on previous exchanges. 
This approach spans commercial platforms (Claude, ChatGPT, Gemini, Perplexity), specialized applications (Google NotebookLM, DeepL Write), developer tools (GitHub Copilot, Cursor, Windsurf) and open-source alternatives (OpenWebUI, LMStudio, JAN, Aider).

To enhance user experience within this restrictive interaction framework, interfaces have implemented various enhancements. 
For example, some interfaces offer limited branching capabilities, allowing users to edit previous messages to create alternative paths, though typically at the cost of erasing the subsequent conversation history to the original message. 
Additional incremental improvements and interface features are summarized in Table~\ref{tab:llm-features}.

\begin{table}[htbp]
    \centering
    \caption{Supplementary Features in LLM Interfaces}
    \label{tab:llm-features}
    \begin{tabular}{|p{0.3\textwidth}|p{0.6\textwidth}|}
        \hline
        \textbf{Feature} & \textbf{Description} \\
        \hline
        Multimodal capabilities & Support for diverse input and output types (text, images, audio, video) \\
        \hline
        Tool integration & Web search, code execution, and protocols (like Model Context Protocol) to interact with external services \\
        \hline
        Separated content stages & Content areas (like Claude artifacts) distinct from conversation flows \\
        \hline
        Project workspaces & Environments where files are automatically added to conversation context \\
        \hline
        Writing style customization & Preset and custom writing styles to control tone, formality, or voice (e.g., concise, explanatory, formal) \\
        \hline
        Profile preferences & Account-wide settings that apply across all conversations \\
        \hline
    \end{tabular}
\end{table}

However, these enhancements focus mainly on extending LLM capabilities rather than addressing interface limitations.
Current designs cannot properly counteract context fragmentation, fail to distinguish between important long-term and merely short-term relevant information, and impose burdensome message management on users. 
For example, users cannot selectively mask irrelevant parts of previous exchanges, dynamically reorder messages, or optimize the context window based on relevance rather than recency. 
Furthermore, it also constrains exploration of problem spaces that would benefit from parallel inquiry paths, particularly for professional contexts where users engage in extended problem-solving.

Recent research has begun addressing these limitations.
Masson et al. (2024) proposed DirectGPT \cite{directgpt}, a system that applies direct manipulation as an additional layer on linear chat interfaces like ChatGPT.
Their approach includes an undo mechanism and a customizable toolbar to store frequently used commands (e.g., replace selected word with synonym) which can be used to update text in-place.
So rather than progressing through continuation of the conversation, they allow users to update the current state continuously with visual feedback for changes.
While DirectGPT demonstrates the potential benefits of moving beyond linear chat interfaces, our work explores a more comprehensive restructuring of the interaction model to address context management, parallel inquiry paths, and dynamic content organization.

\subsection{Human-Centered Design Process} \label{ch:hcd}
The human-centered design (HCD) process provides a methodological framework for developing interactive systems that prioritize user needs. Standardized in DIN EN ISO 9241-210 \cite{din2010iso}, HCD diverges from linear development approaches like the waterfall model by Royce \cite{royce1987managing} through its iterative, user-focused process.

\begin{figure}
    \centering
    \includegraphics[width=0.5\linewidth]{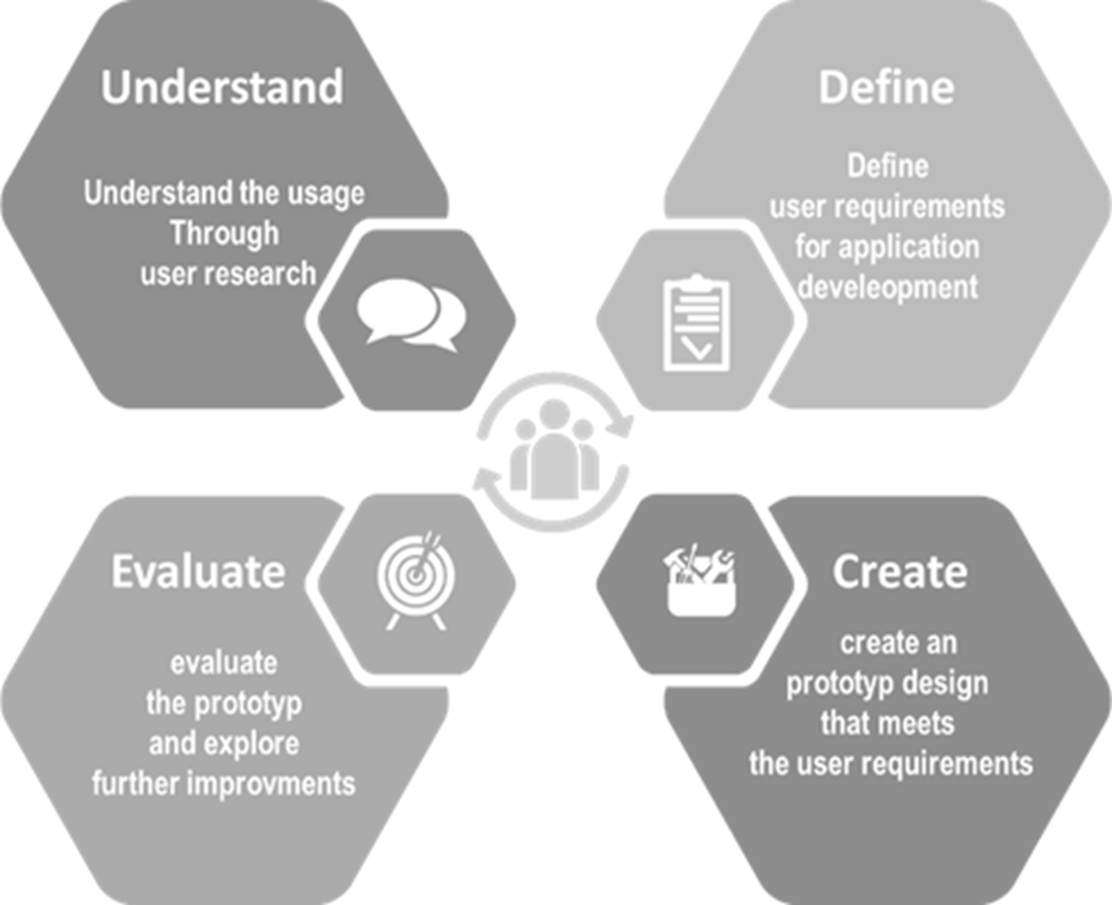}
    \caption{Phases of the human-centered design process (own illustration)}
    \label{fig:hcd}
\end{figure}

The HCD process comprises the four phases: understanding the usage context, defining the usage requirements, drafting design solutions and evaluating the design solutions against these requirements. These phases are run through iteratively until an optimal result is achieved. The process is visualized in Figure \ref{fig:hcd}.

For our work, HCD offers a structured approach for user-centric interface development for LLM, whereby this paper focuses on the initial conceptualization phase of this process. The HCD paradigm was choosen for our project as previous research shows that development following these principles can lead to various advantages. 
Human-centered design has been shown to improve usability, user satisfaction, and task performance, particularly in complex socio-technical systems. 
It emphasizes iterative evaluation with real users, thereby reducing design errors and mismatches between system behavior and user expectations \cite{norman1986cognitive, boy2011introduction}. 
Moreover, HCD enhances situation awareness, adaptability to varied user populations, and alignment with cognitive processes—factors that are critical in the design of AI-driven interfaces such as those involving large language models (LLMs) \cite{boy2011introduction, endsley1996situation}. 
By grounding interface design in the real needs and contexts of users, HCD contributes not only to system efficiency but also to safety and long-term system acceptance \cite{billings1991human, hollnagel2005joint}.
\section{Workshop Structure}
To ensure user participation in the development of a new LLM interface, we have organized a design thinking workshop titled \href{https://events.hifis.net/event/1741/contributions/14074/}{Breaking the Chat Barrier} at the deRSE2025 in Karlsruhe.  
Design thinking, utilizing the logic of the HCD process, is an iterative process of empathizing with users, defining their core challenges, ideating solutions, prototyping those ideas, and testing them for feedback \cite{dorst2011core}. The aim of the workshop was to gather use cases of LLMs and to identify the challenges and barriers the users face.  Based on these insights, we collaboratively identified requirements for a dynamic chat interface and developed different visualization concepts. 
The workshop was held on the first day of the conference and lasted 60 minutes. The workshop materials are available on the \href{https://events.hifis.net/event/1741/contributions/14074/attachments/3430/7158/deRSE25\_LLM\_UI_Workshop\_24.02.25\_Frank\_Lund.pdf}{HiFis conference portal}.

Since deRSE is attended by scientists from diverse disciplines who work with research software -- whether as programmers or in other roles -- this audience provides two key advantages for our workshop: 
First, most attendees are likely already using LLMs in their daily or private work and therefore bring firsthand experience with this technology. 
Second, their general programming background might be helpful to imagine improvements for LLM interfaces. 
Because the workshop was open to all conference participants and registration was voluntary, we cannot report any demographic information about the attendees.

The workshop consisted of an introductory presentation and group work, which in turn was divided into several tasks. 
The introduction should bring all participants to a similar level, regardless of their prior knowledge. 
To this end, we gave a brief introduction to the architecture of LLMs, human-computer interaction frameworks and the HCD process. 
In addition, interfaces of existing LLMs were presented, and their limitations were demonstrated using the user story of a persona, the fictitious programmer Frank. 
For further details of the introduction, please refer to the slides linked above.

The group work consisted of four different tasks, each of which is reproduced here in the wording used in our presentation.
At the beginning of the workshop, we asked the participants \textbf{"how you use LLMs"} and collected their answers on a whiteboard.
No restrictions were placed on the type of answers, so participants could describe both business and personal use cases.
In a follow-up question, we ask the participants to \textbf{"note what you like and dislike about working with LLMs"} on another card.

As mentioned above, screenshots of the UIs of various LLMs were shown in our introductory presentation, whereupon we asked the participants \textbf{"what do these [user] interfaces have in common"}? 
This question served as a ramp for the last and most important group work, in which the participants were asked to create visualizations for new UIs based on the guiding question \textbf{"How to overcome the limitations of chat based LLM interfaces to enable new potentials?"}. 
This technique is also called a \textit{How-Might-We-question} and is often utilized as a design thinking method to set a user centered mindset for brainstorming and prototyping. 
The participants were divided into seven small groups of 4 to 6 people each to visualize the UI using a flipchart. 
They were given 15 minutes for this task in total, 5 minutes for an initial discussion and 10 for the visualization process itself. The design solutions were then presented and discussed by the entire group, bringing the workshop to a close.
\section{Results} \label{ch:results}

\subsection{Evaluation of the Workshop Results}
For the evaluation of our workshop outputs, we first transcribed every written response --- from the whiteboard notes and flip chart cards - as verbatims into a single document. The documentation was done manually and without software for quantitative analysis, as the dataset was moderate in size. 
We preserved participants' original wording to avoid introducing interpretive bias. 
Next, we conducted a two-stage, inductive thematic grouping consistent with conventional qualitative content analysis (i.e., deriving categories directly from the data without imposing preconceived codes) as described by \cite{hsiuThreeApproaches}. 
In the first stage, we reviewed each item and clustered those whose meaning was unmistakably identical, grouping them under provisional labels. 
In the second stage, we assigned these provisional clusters to higher-level categories, again merging only when semantic overlap was clear without interpreting latent meanings. 
Any responses that could not be confidently aligned with other items remained as standalone categories.\\

We followed the following rules for creating our response tables:

\begin{itemize}
    \item \textbf{Ordering within categories:} Entries are listed in descending order by frequency of mention; ties are ordered randomly.
    
    \item \textbf{Whole-response assignment:} If a single response addressed multiple issues, it was not split across categories; instead, the response was preserved intact and assigned to the single most fitting category.
    
    \item \textbf{Alternative category notes:} Where an equally justifiable alternative category existed, we indicated it in italics within parentheses immediately after the entry.

    \item \textbf{Table conventions:} Column headers explain each category and display the number of responses. Table cells contain the participants’ verbatim responses, so their specificity varies according to how each individual chose to express their point.
\end{itemize}

This conservative, semantic-level approach parallels the affinity-diagramming method widely used in human-computer interaction research to organize unstructured user data into theme clusters for design insights \cite{holtzblattContext}. 
By staying very close to the text, our summary reflects the diversity of participants' answers while reducing redundancy and highlighting the most salient themes.

\newpage
\subsection{Participants' Experience with LLMs}
Scientists' uses of LLMs fell into two main domains: \textit{coding} (22 mentions) and \textit{human-language tasks} (34 mentions). Within coding, the bulk of responses (14 mentions) described the coding process itself—writing snippets for apps and web interfaces, auto-completing functions, checking syntax on the fly, and even refactoring existing modules—while a smaller set (8 mentions) focused on code assisting, such as generating documentation, explaining library functions, and troubleshooting error messages.

In the realm of language, participants’ needs split across five distinct uses. Seven attendees relied on the model for narrow, goal-oriented text creation—drafting emails, turning bullet points into prose, polishing paper titles, or even assembling a grocery list—while five embraced its wide-creative potential by brainstorming workout plans, interview questions, or exam prompts. Eleven mentions of text improvement highlighted LLMs’ strength in proofreading, clarity edits, and stylistic rewrites, and another seven responses tapped into their ability to unpack and summarize complex material, answer detailed questions, or simplify technical jargon. Finally, four miscellaneous uses ranged from practicing a foreign language in tandem to extracting text via OCR for literature overviews. 

Together, these patterns underscore that scientists lean on LLMs both as practical coding companions and as versatile language partners—able to deliver precise, task-driven outputs or to spark broader creative and analytical support.

\newpage
\thispagestyle{empty}
\begin{landscape}

\begin{longtable}{|p{3cm}|p{2.5cm}|p{3cm}|p{3cm}|p{3cm}|p{3.5cm}|p{3.5cm}|}
\caption{Responses to “How you use LLMs”}\\
\hline
\multicolumn{2}{|c|}{\textbf{Coding (22)}} & \multicolumn{5}{c|}{\textbf{Human language (34)}} \\
\hline
\multicolumn{1}{|c|}{\textbf{Coding process}} & 
\multicolumn{1}{c|}{\textbf{Code assisting}} & 
\multicolumn{1}{c|}{\textbf{Text creation -}} & 
\multicolumn{1}{c|}{\textbf{Text creation -}} & 
\multicolumn{1}{c|}{\textbf{Text improvement}} & 
\multicolumn{1}{c|}{\textbf{Text understanding}} & 
\multicolumn{1}{c|}{\textbf{Text miscellaneous}} \\
\multicolumn{1}{|c|}{\textbf{(14)}} &
\multicolumn{1}{c|}{\textbf{(8)}} &
\multicolumn{1}{c|}{\textbf{goal narrow / task}} & 
\multicolumn{1}{c|}{\textbf{goal wide /}} & 
\multicolumn{1}{c|}{\textbf{(11)}} &
\multicolumn{1}{c|}{\textbf{(7)}} &
\multicolumn{1}{c|}{\textbf{(4)}}
\\
\multicolumn{1}{|c|}{} &
\multicolumn{1}{c|}{} &
\multicolumn{1}{c|}{\textbf{oriented (7)}} & 
\multicolumn{1}{c|}{\textbf{creative oriented (5)}} & 
\multicolumn{1}{c|}{} &
\multicolumn{1}{c|}{} &
\multicolumn{1}{c|}{} 
\\
\hline
\endhead
\hline
\endfoot
Writing code (6x) (e.g., small apps, web interfaces, config files) & Writing documentation (2x) & Write text based on bullet points / keywords (2x) & Idea generation (2x) & Quality improvement (6x): (e.g., proof-reading, rewriting) & Summarization (4x) & Tandem partner for language learning \\
\hline
Generate code snippet (5x) & Code explanation (2x) & Writing emails (2x) & Organizing workout schedule & Rephrasing (4x): & Provides context to certain topics & Gathering literature (paper, references) and brief summary for new topics \\
\hline
Code completion \& generation & Finding \& understanding library function (2x) & Job application writing & Interview preparation & Make suggestions on contents of a document & Explanation of research topics & Text recognition from images \\
\hline
Checking syntax while programming & Solutions for resolving programming issues & Generating / tweaking of paper titles & Homework creation / question creations for exams & & Answer questions about content in our textbook (e.g., what is the CEO of a hospital responsible for) & The investigation of the analytical solution of the local strain around nanoparticles in incompressible elastic material \\
\hline
Code refactoring & Programming help \textit{(could also belong to category "Coding process")} & Preparing grocery list & & & Simplifying text for better understanding \textit{(could also belong to category "Text improvement")} & \\
\hline
\end{longtable}
\end{landscape}

Participants highlighted six ways in which LLMs enhance their workflows. First, on the \textit{usability} front (3 mentions), users liked the models' simplicity and conversational feel -- “easy to use” (2×), straightforward setup even for local deployment, and an interactive, dialogue-like interface that feels more natural than form-based tools.

Closely related were comments about \textit{user experience} (4): several appreciated that the AI is “always in a helpful mood,” and that engaging with it requires minimal mental overhead (“we don’t need to think a lot”). Others noted how it unlocks new workflow possibilities and spares them from tedious tasks—one participant admitted, “I don’t like writing documentation, so AI is really great there.”

\textit{Efficiency and productivity}  were the most frequently cited advantages (16 mentions). Participants particularly emphasized time savings (9 mentions), but also highlighted the model’s attention to detail, the ability to receive multiple alternative suggestions in a single response, and generally reliable accuracy.

Next, on the \textit{quality of outputs} (9), attendees positively mentioned how the model consistently produces polished, ready-to-use artifacts. Two participants specifically noted that it “improves text quality,” while another pair singled out its “professional communication style.” Beyond prose, users found the LLM “works reasonably” for basic code refactoring, debugging, documentation, and translation tasks. The participants also liked its nuanced “attention to detail and multiple suggestions,” described outputs as “mostly accurate,” and highlighted its proficiency in correction, summarization, and stylistic edits—especially for concise summarization tasks.

On the \textit{functional capabilities} side (8 mentions), participants noted that the LLM effectively follows common conventions and standards, retains instructions across multiple turns, and often produces useful output even from minimal prompts. Participants found the model to be adept at answering simple, direct questions and appreciated its depth of topic knowledge as well as its syntactic accuracy.

Please note, that we differentiate these two categories as follows:  while \textbf{quality of outputs}, captures how well each individual response is crafted (e.g. accuracy, style, level of detail), \textbf{functional capabilities}, by contrast, describe the broader classes of tasks the model can solve reliably rather than the fine-grained quality of any single answer.

Finally, in \textit{knowledge and learning support} (3), users pointed out how LLMs lower the barrier to exploring unfamiliar fields—providing broad coverage of topics, making it easy to dive into new areas, and even decoding cryptic error messages in R or other languages.

Together, these highlights show that scientists value LLMs not only for what they can do, but also for how seamlessly they integrate into both coding and research tasks—streamlining work while maintaining an approachable, user-friendly experience.

\newpage
\thispagestyle{empty}
\begin{landscape}
{\small
\begin{longtable}{|p{3cm}|p{3cm}|p{3cm}|p{4cm}|p{4cm}|p{3cm}|}
\caption{Responses to “Note what you like about working with LLMs”}
\label{tab:llm-likes-dislikes}\\
\hline
\multicolumn{1}{|c|}{\textbf{Usability}} & 
\multicolumn{1}{|c|}{\textbf{User experience (UX)}} & 
\multicolumn{1}{c|}{\textbf{Efficiency \&}} & 
\multicolumn{1}{c|}{\textbf{Quality of outputs}} & 
\multicolumn{1}{c|}{\textbf{Functional capabilities of}} & 
\multicolumn{1}{c|}{\textbf{Knowledge \&}} 
\\
\multicolumn{1}{|c|}{\textbf{(3)}} &
\multicolumn{1}{|c|}{\textbf{(4)}} &
\multicolumn{1}{|c|}{\textbf{productivity}} &
\multicolumn{1}{|c|}{\textbf{("How good and usable is}} &
\multicolumn{1}{|c|}{\textbf{LLMs ("What kinds of}} &
\multicolumn{1}{|c|}{\textbf{learning support}} 
\\
 &
 &
\multicolumn{1}{|c|}{\textbf{(16)}} &
\multicolumn{1}{|c|}{\textbf{the actual generated}} &
\multicolumn{1}{|c|}{\textbf{tasks can the LLM perform}} &
\multicolumn{1}{|c|}{\textbf{(3)}}
\\
 &
 &
 &
\multicolumn{1}{|c|}{\textbf{output?") (9)}} &
\multicolumn{1}{|c|}{\textbf{effectively?") (8)}} &
\\
\hline
\endhead
\hline
\endfoot

Easy to use (2x) & Always in a helpful mood & Time savings (9) & Improves text quality (2x) & Follows common conventions and standards programming (3x) & Wide topic coverage and reduced effort for learning new topics \\
\hline

Ease of setup \& local usage & We [the user] don't need to think a lot & Fast (5x) & Professional communication style (2x) & GPT-Projects: Remembering the instructions (No need to instruct again for same task) & Easy to dig into new topics \\
\hline

Interactive "conversation"-like & New possibilities due to workflow changes & Efficiency & Works reasonably for simple forms of code refactoring, debugging, documentation, translation & Even with a small prompt, LLMs can usually expand a lot which is really good & Explains cryptic R error messages very well \\
\hline

& I don't like writing documentation, so AI is really helpful in overcoming the burden & Sometimes increased productivity & Attention to detail and providing multiple suggestions & Very useful for simple questions with direct answers & \\
\hline

& & & Mostly accurate & Deep knowledge on the topic and syntax & \\
\hline

& & & Good at correction, summarizing, and style changes of text & Builds boilerplate code rapidly, recommends libraries and gives corrections for specific cases & \\
\hline

& & & Good for summarization [of] task & & \\
\hline

\end{longtable}
}
\end{landscape}

Participants surfaced a range of dissatisfactions with LLMs that fall into five main areas.
First, \textit{functional capabilities and usability} (15 mentions) topped the list. Several users found outputs “wordy” or “repetitive” (3×), and many noted that the model sometimes “ignores coding instructions” (2×) or “doesn’t understand the intention of the question.” Others mentioned persistence errors—losing context mid-conversation—and pointed out that some tools “only work well with widely used software,” limiting their flexibility. Only one person criticized a lack of usability while mentioning that "copy / paste by hand via web interface is not very practical".

Closely related was \textit{output reliability and accuracy} (13). “Hallucinations” (4×) and “faulty code” (4×) were recurring pain points, alongside general “factual mistakes” (3×) and the sense that “the solution suggested is wrong.” Even minor inaccuracies eroded trust, reinforcing that users need dependable, verifiable results.

With growing awareness of data practices, \textit{privacy, environmental, and social concerns} also emerged (13). Three mentions each for “data safety” and worries about “exploitation of labor and knowledge for training” signaled unease about how user inputs and large-scale data harvesting occur. Some participants even questioned “companies using data for nefarious reasons,” underscoring ethical considerations beyond mere tool performance.

Under \textit{explainability and transparency} (5), attendees criticized the opacity of model internals: “lack of transparency” and “lack of provenance” each featured twice, and several noted it’s “unclear what model is the best for a use case.” Without clear insight into how decisions are made, users felt handicapped when evaluating or debugging the LLM’s reasoning.

Finally, \textit{disillusionment and saturation} (9) captured deeper, more subjective concerns: five participants described a “loss of cognitive engagement” (“if I always rely on AI, I learn less”), while others warned of “false security,” “AI dependency,” and fear that “people become dumb” by outsourcing thinking. One user even remarked that they “can identify LLM-generated text—and its ever-present style—everywhere,” hinting at early signs of fatigue and frustration.

Together, these critiques illuminate not just technical shortcomings, but broader worries about trust, ethics, and the evolving role of AI in society.
\newpage
\begin{landscape}
{\small
\begin{longtable}{|p{3.5cm}|p{3.5cm}|p{4cm}|p{3.5cm}|p{3.5cm}|}
\caption{Responses to “Note what you dislike about working with LLMs”}\\
\hline
\multicolumn{1}{|c|}{\textbf{Lack of functional}} & 
\multicolumn{1}{c|}{\textbf{Output reliability \&}} & 
\multicolumn{1}{c|}{\textbf{Data privacy \&}} & 
\multicolumn{1}{c|}{\textbf{Explainability \&}} & 
\multicolumn{1}{c|}{\textbf{Disillusionment /}} 
\\
\multicolumn{1}{|c|}{\textbf{capabilities (15) /}} & 
\multicolumn{1}{c|}{\textbf{accuracy}} & 
\multicolumn{1}{c|}{\textbf{environmental \& social}} & 
\multicolumn{1}{c|}{\textbf{transparency}} & 
\multicolumn{1}{c|}{\textbf{saturation}} 
\\
\multicolumn{1}{|c|}{\textbf{usability (1) \footnotemark}} & 
\multicolumn{1}{c|}{\textbf{(13)}} & 
\multicolumn{1}{c|}{\textbf{concerns (13)}} & 
\multicolumn{1}{c|}{\textbf{(5)}} & 
\multicolumn{1}{c|}{\textbf{(9)}} 
\\
\hline
\endhead
\hline
\endfoot

Wordy / repetitive (3x) & Hallucinations (4x) & Data safety (3x) & Lack of transparency (2x) & Loss of cognitive engagement (5x) (e.g., if I am not mindful, I do not learn, encourages use without understanding) \\
\hline

Persistence errors (e.g., losing or forgetting information after long discussions) (2x) & Faulty code (4x) & Privacy (3x) & Unclear what model is the best for a use case (e.g., not specified on Huggingface leaderboard) & False security \\
\hline

Ignoring coding instructions (2x) & Factual mistakes (3x) & Copyright issues (2x) & Lack of explainability & AI dependency concerns \\
\hline

Only works well with widely used software & Not always accurate (2x) & Companies using data for nefarious reasons & Lack of provenance & People becoming dumb, using it everywhere \\
\hline

Doesn't understand the intention of the question & The solution suggested from LLM is wrong & Exploitation of labour \& knowledge for training AI-System (great tech, but a system based not on openness + quality, etc.) & & I can identify LLM-generated text and it's everywhere \\
\hline

Settings dependent on LLMs & Sometimes gives wrong conclusion & Waste of natural resources & & \\
\hline

Might do incorrect debugging & Verification of output needed & Too much hardware required to train LLMs \textit{(could also belong to category "Lack of functional capabilities")} & & \\
\hline

Tendency to please (rather give an incorrect answer than state that something doesn't work) & Often wrong (even so slightly) & It is often visible that the LLM was mainly trained around English language materials \textit{(could also belong to category "Lack of functional capabilities")} & & \\
\hline

Finetuning didn't work for us & & & & \\
\hline

Knowledge required on how to prompt efficiently & & & & \\
\hline

Uncensoring is annoying \textit{(could also belong to the category "Data privacy \& environmental \& social concerns")} & & & & \\
\hline

Copy / paste by hand via web interface is not very practical & & & & \\
\hline

\end{longtable}
}

\footnotetext{Whilst \textit{usability} is its own category, we have included the only answers to this area in the same column as the \textit{functional capabilities} to ensure better readability of the whole table.}
\end{landscape}

\newpage
\subsection{Participant-Designed UI Solutions}
For the final task of the workshop, we asked participants to sketch and present new concepts for LLM interaction forms. All seven solutions are described below.

\begin{figure}[!thp]
  \centering
  \begin{minipage}[b]{0.45\linewidth}
    \includegraphics[width=\linewidth]{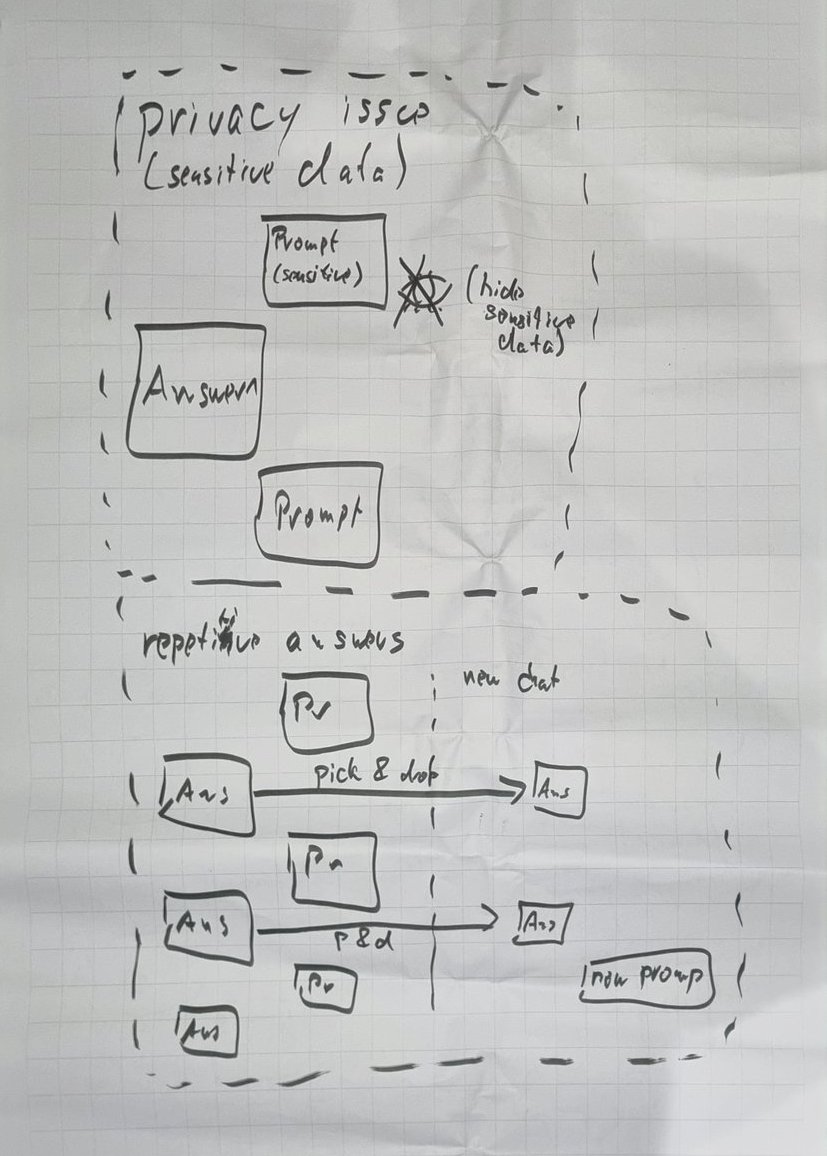}
    \caption{UI Solution 1}
    \label{fig:sol_01}
  \end{minipage}
  \hfill
  \begin{minipage}[b]{0.45\linewidth}
    \includegraphics[width=\linewidth]{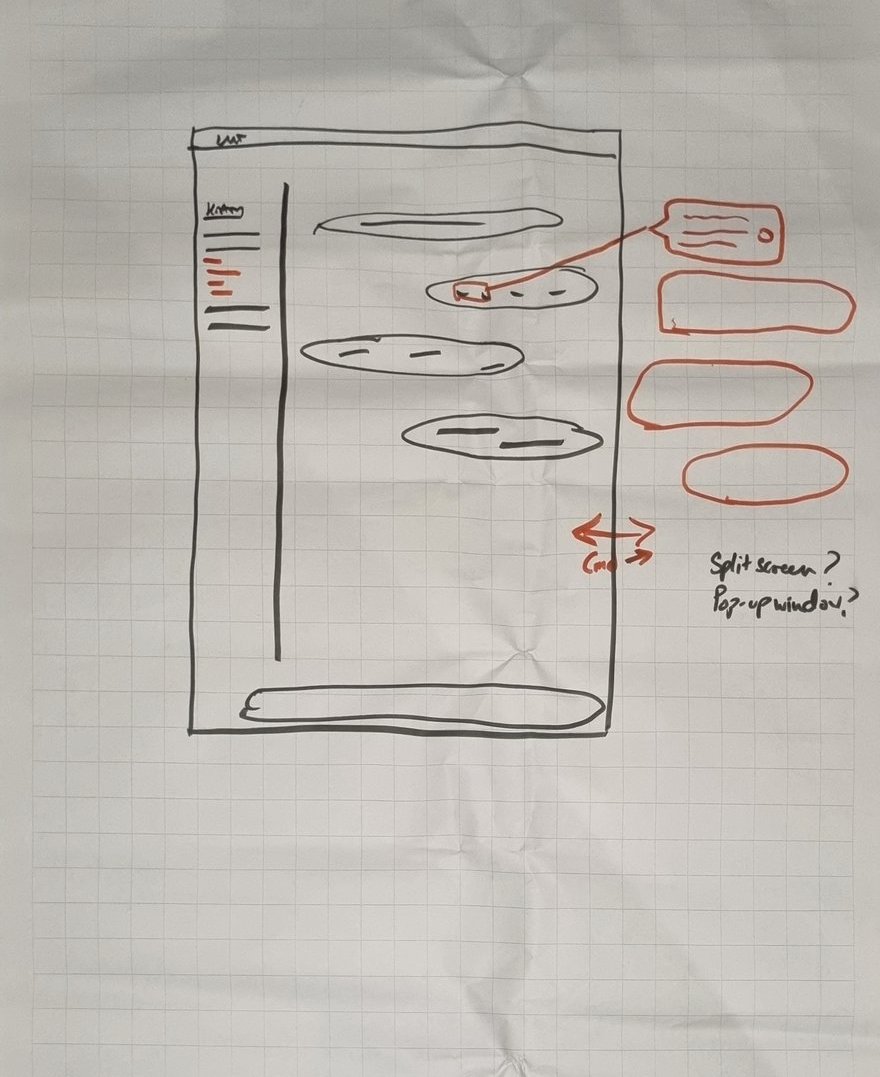}
    \caption{UI Solution 2}
    \label{fig:sol_02}
  \end{minipage}
\end{figure}

\noindent
\textbf{UI Solution 1} (see Figure \ref{fig:sol_01}).
This concept introduces a horizontal panel-based chat interface, where each panel represents a distinct chat. Users can branch their conversation by dragging only the answers they find useful into a new panel to the right, creating a new chat context built from selected outputs. Repetitive or unsatisfactory answers are left behind, allowing users to curate and continue with a “clean” conversation history. The upper part of the graphic, which sketches a privacy function for hiding sensitive data, is not considered here as developing new privacy features is out of scope for our current interaction concepts.\\

\noindent
\textbf{UI Solution 2} (see Figure \ref{fig:sol_02}).
This concept explores an interface where hovering over individual prompts or answers allows users to create branches, opening a secondary conversation as an overlay or pop-up directly next to the original chat bubble. These sub-branches function like tooltips, letting users explore alternative paths or clarifications without leaving the main thread. Expanded sub-conversations can be collapsed again to maintain clarity and reduce clutter. On the far right, a  timeline visually summarizes all ongoing and past LLM conversations.

\newpage
\begin{figure}[!thp]
  \centering
  \begin{minipage}[b]{0.45\linewidth}
    \includegraphics[width=\linewidth]{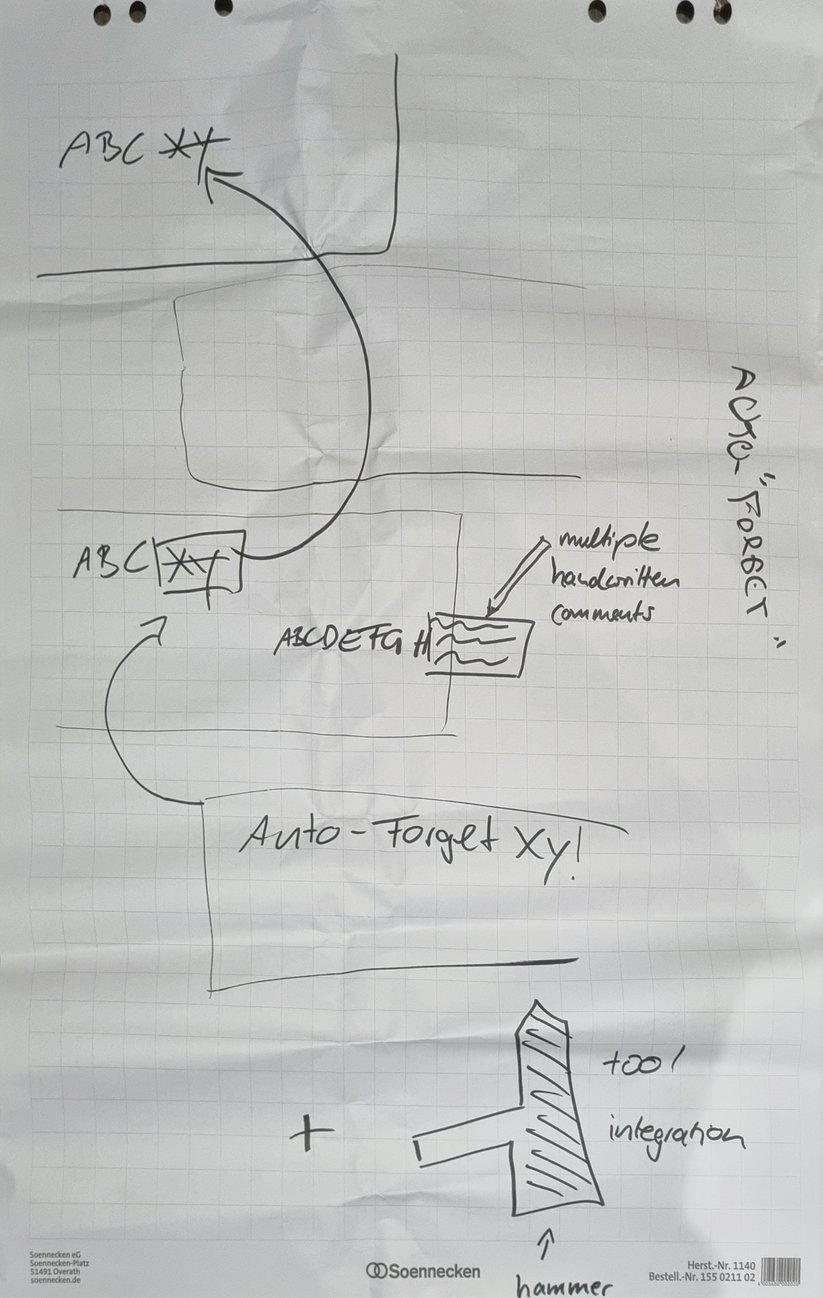}
    \caption{UI Solution 3}
    \label{fig:sol_03}
  \end{minipage}
  \hfill
  \begin{minipage}[b]{0.45\linewidth}
    \includegraphics[width=\linewidth]{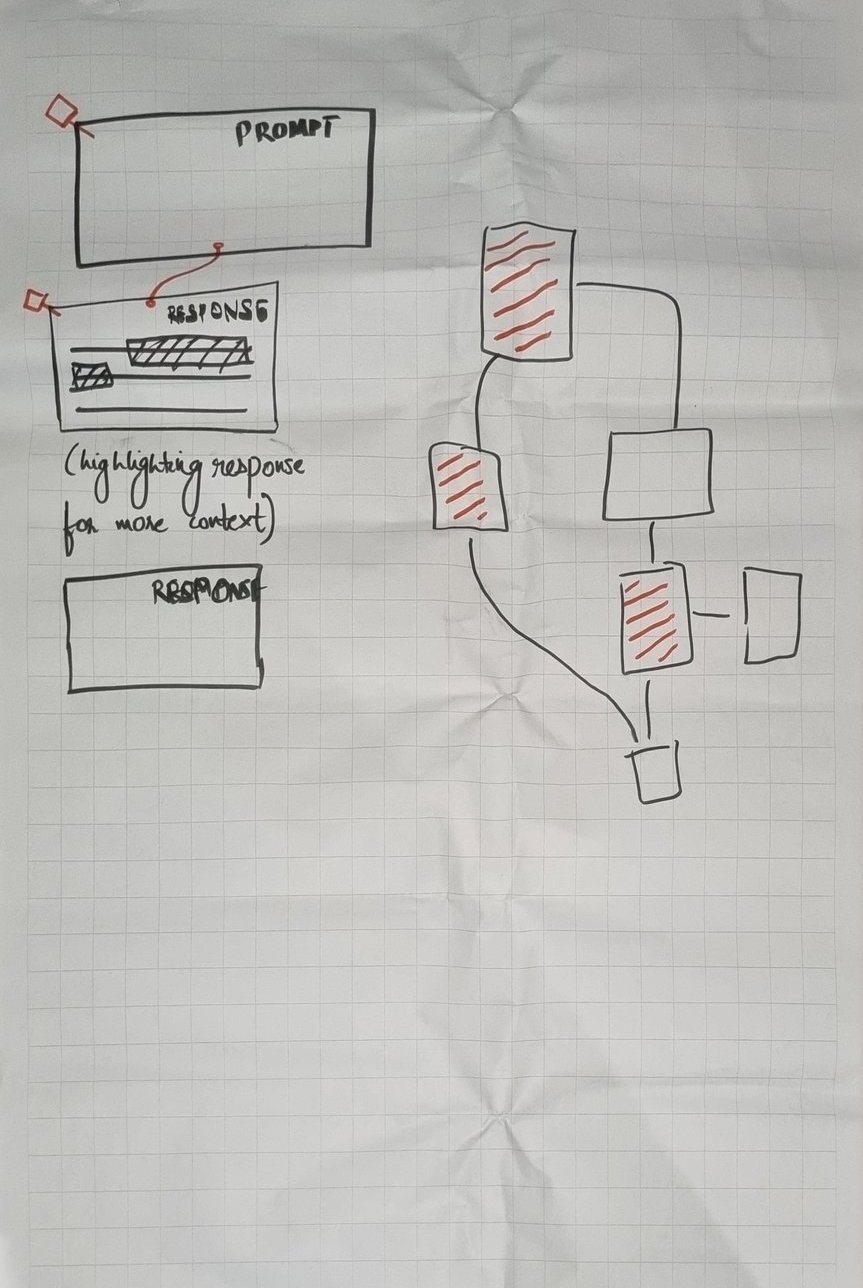}
    \caption{UI Solution 4}
    \label{fig:sol_04}
  \end{minipage}
\end{figure}

\noindent
\textbf{UI Solution 3} (see Figure \ref{fig:sol_03}).
This concept introduces an “auto-forget” function that allows users to explicitly instruct the LLM to forget specific elements or sections of the conversation, for example by typing a prompt like “auto-forget [content].” The system then automatically identifies and removes all mentions or context related to that element from the ongoing conversation, ensuring subsequent outputs do not reference it. The action is visualized with a hammer icon, reinforcing the idea of forcefully removing information.\\

\noindent
\textbf{UI Solution 4} (see Figure \ref{fig:sol_04}).
This concept combines an LLM chat interface on the left with a branching visualization on the right, showing how the conversation structure evolves. Users can pin entire messages or highlight specific parts of a response to indicate their importance. These highlighted or pinned elements are intended to have increased influence on subsequent LLM outputs, effectively weighting them more in the model’s context. The visualization on the right makes it easy to track how branches form whenever a new response is generated based on different highlights.

\newpage
\begin{figure}[!thp]
  \centering
  \begin{minipage}[b]{0.45\linewidth}
    \includegraphics[width=\linewidth]{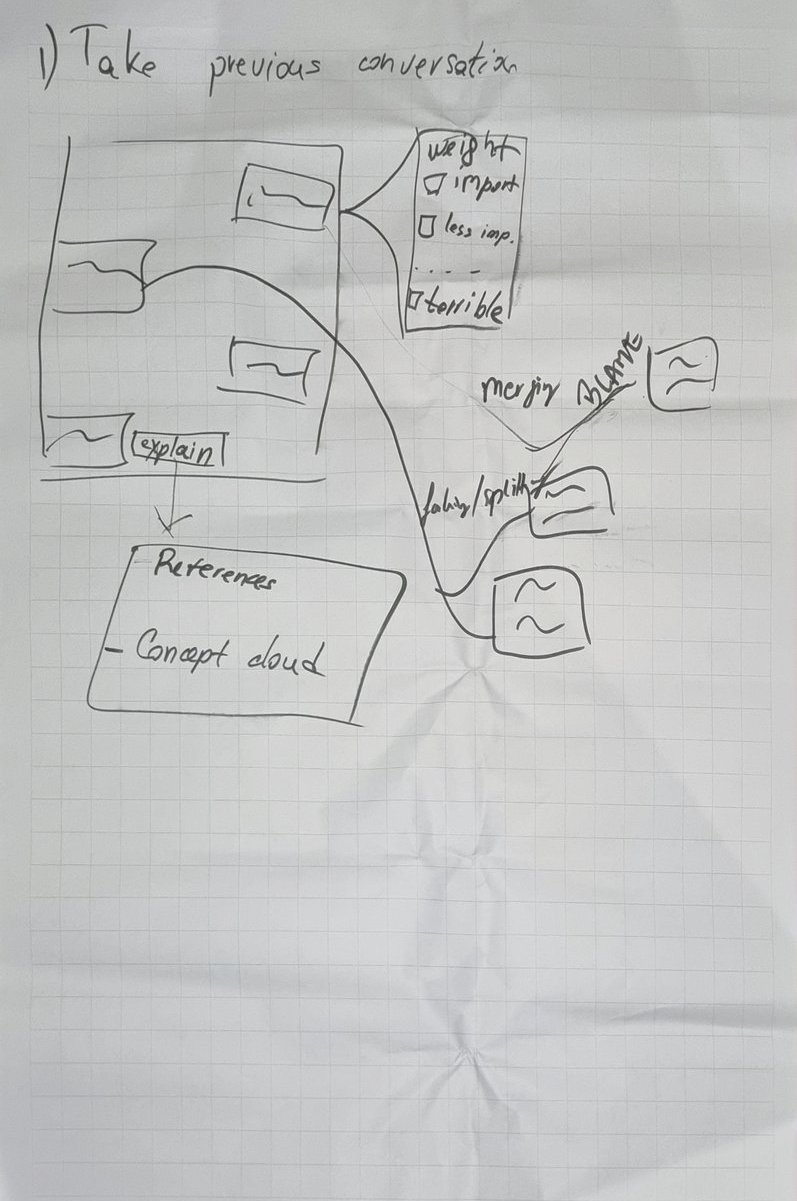}
    \caption{UI Solution 5}
    \label{fig:sol_05}
  \end{minipage}
  \hfill
  \begin{minipage}[b]{0.45\linewidth}
    \includegraphics[width=\linewidth]{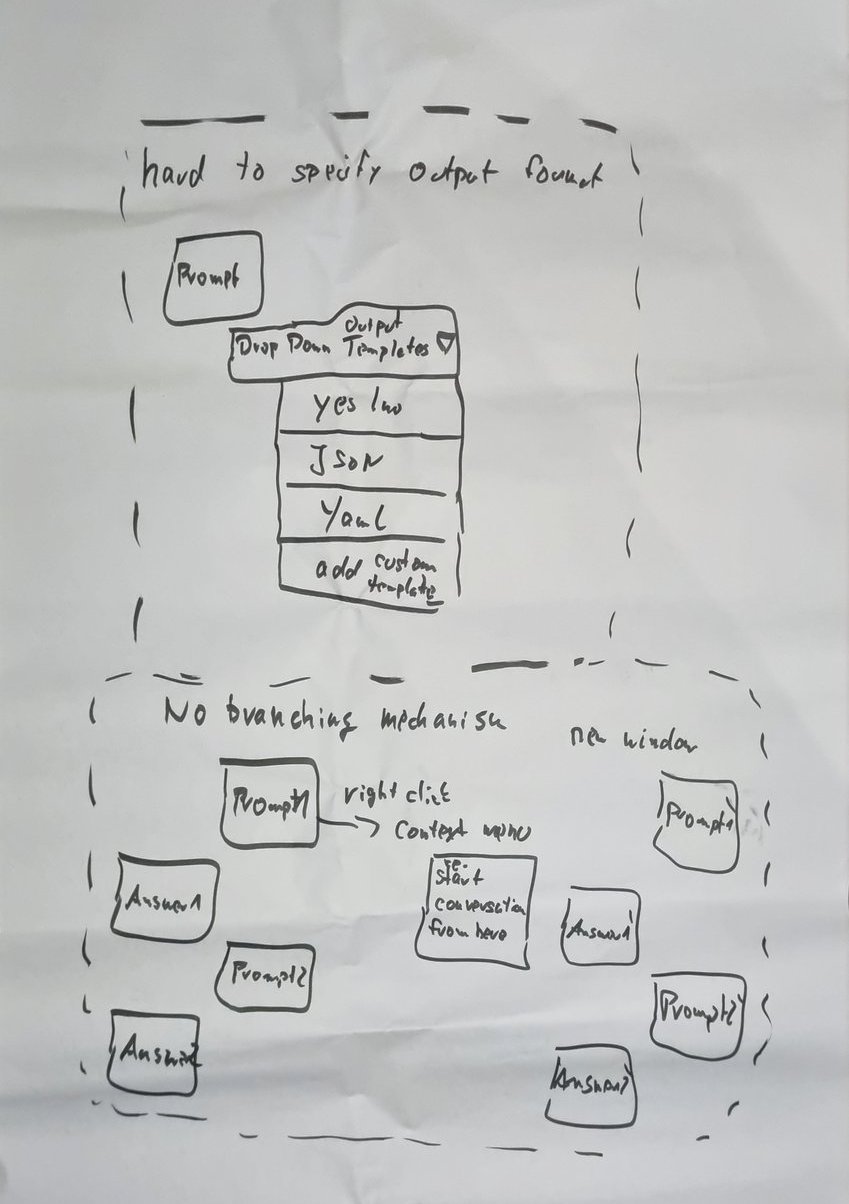}
    \caption{UI Solution 6}
    \label{fig:sol_06}
  \end{minipage}
\end{figure}

\noindent
\textbf{UI Solution 5} (see Figure \ref{fig:sol_05}).
This concept shares similarities with the previous branching and highlighting visualisation in figure 5 but introduces an explicit weighting mechanism: users can click on any prompt or message and assign it a value on a scale from “important” to “unimportant” or even “terrible.” These weights directly influence how much impact each part of the conversation has on subsequent LLM responses. Additionally, users can merge or split conversation threads, and there is a section for references or concept clouds to provide supporting information or context. In contrast to the previous concept - where the chat and the branching visualization are separate - here they are combined and can be displayed dynamically when hovering over the respective prompts.\\

\noindent
\textbf{UI Solution 6} (see Figure \ref{fig:sol_06}).
This concept tries to solve two problems: the difficulty of specifying the desired output format, and the lack of flexible conversation branching. These issues are addressed in the upper and lower panels, respectively. For output formatting (upper panel), users are given a dropdown menu alongside the prompt, allowing them to easily select or define how the LLM’s response should be structured (for example, as JSON, YAML, or with a custom template), rather than relying on prompt engineering. For branching, shown in the lower panel , users can right-click on any previous prompt or answer to open a context menu and choose “start conversation from here,” which adds a new conversation thread to the right side of the UI— similar to concept number one. 

\newpage
\begin{figure}
    \centering
    \includegraphics[width=0.45\linewidth]{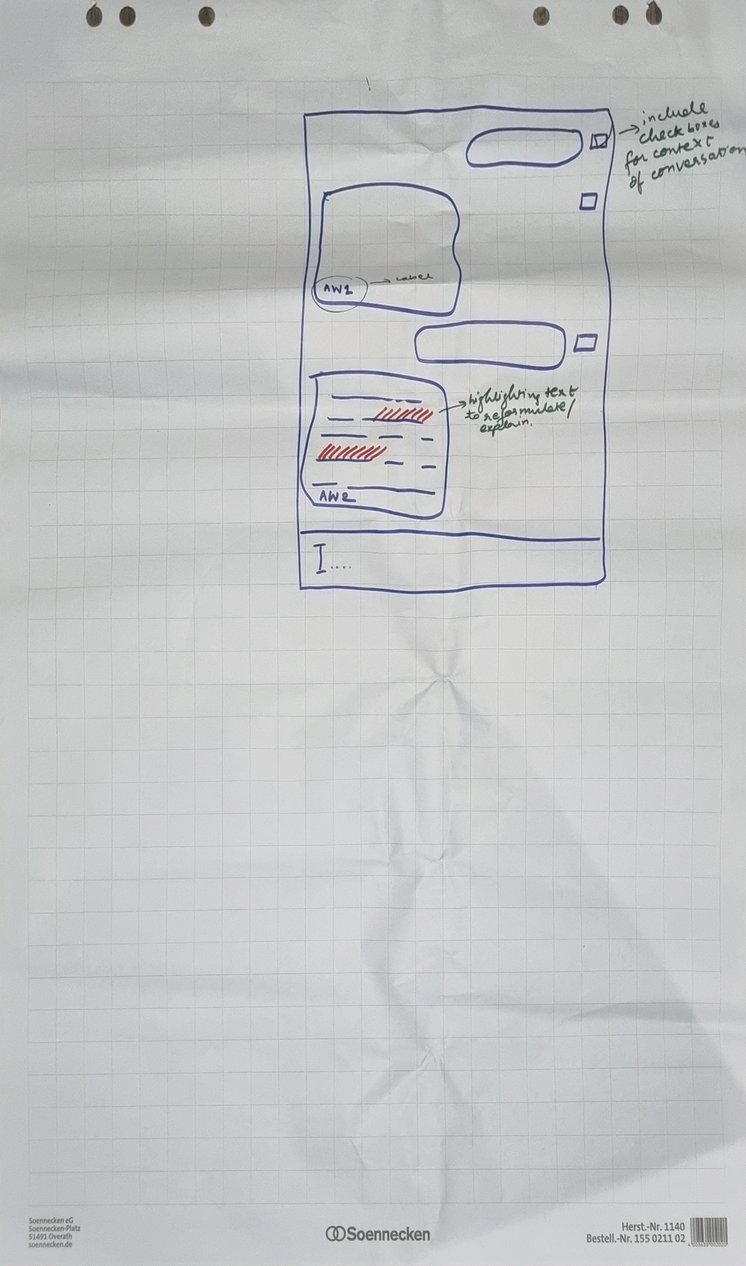}
    \caption{UI Solution 7}
    \label{fig:enter-label}
    \label{fig:sol_07}
\end{figure}

\noindent
\textbf{UI Solution 7} (see Figure \ref{fig:sol_07}).
This concept enables users to flexibly manage the context used by the LLM in generating responses. Each message in the conversation history has a checkbox, allowing users to include or exclude specific messages from the active context. Additionally, users can highlight individual words or sections—even in previous prompts or responses—to request reformulation, clarification ot terms. When changes are made, outputs are regenerated to reflect the updated selections.
\section{Concept}

\subsection{Dynamic Whiteboard UI}
Driven by the perception of unrealized potential in the current state of UIs of LLMs, the authors of this paper have created their own UI concept previous to the Workshop at deRSE25. 
This concept was awarded the AI-HUB@LMU Prize for the Most Innovative AI-based Research Project in 2024. 
In this chapter, we present this concept, explain its functionality and show how its further development will benefit from the results of the design thinking workshop.

As mentioned in the prior chapter, current LLM are hampered by their rigid, chronological chat interfaces. 
In such systems, conversation unfolds chronologically, closely mimicking a dialogue between two parties. 
While this model is intuitive for basic exchanges, it imposes constraints on more complex tasks. 
As a result, users often struggle with cluttered, unfocused contexts, which can diminish the precision and usefulness of LLM outputs.

\begin{figure}[ht]
    \centering
    \includegraphics[width=0.85\linewidth]{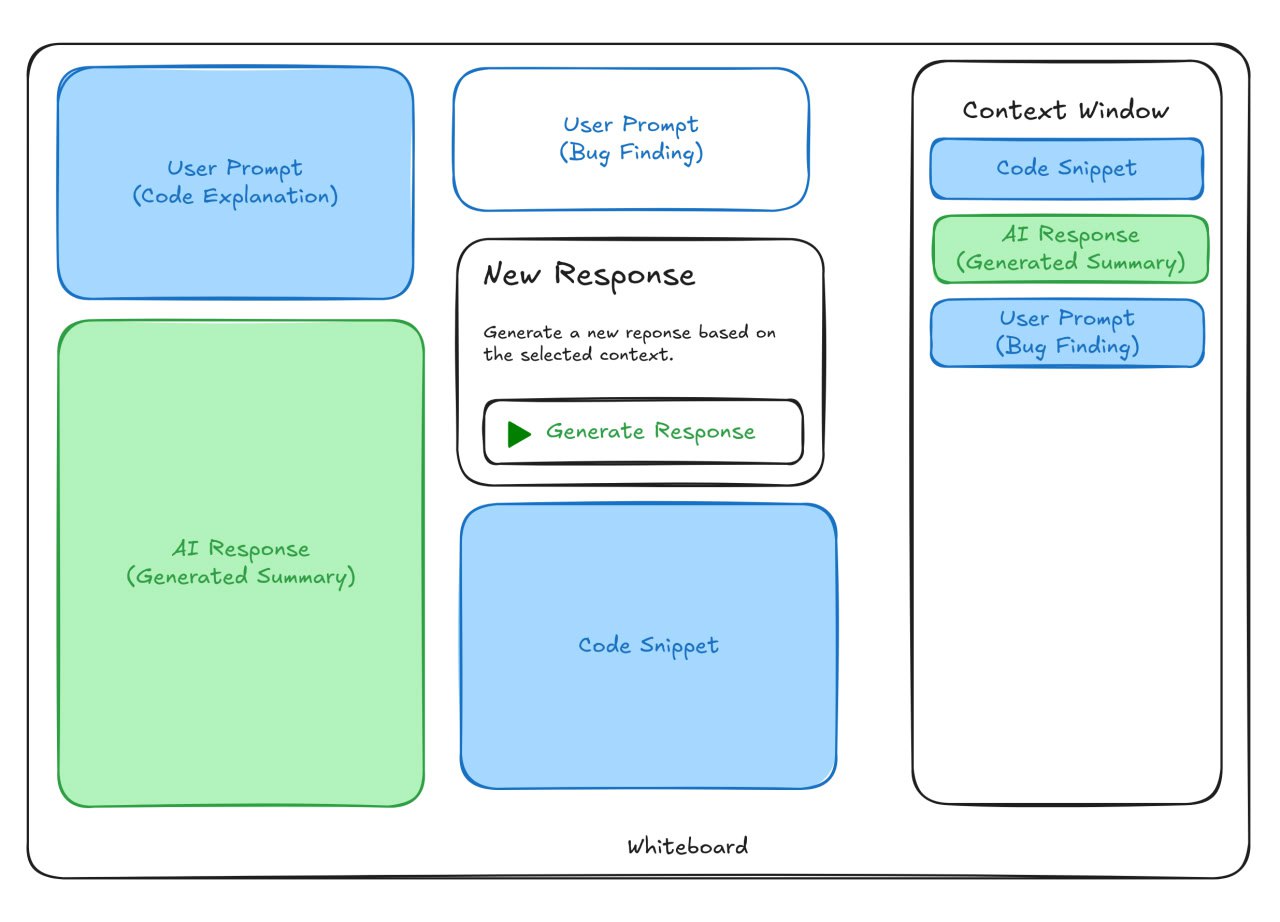}
    \caption{Concept for a dynamic whiteboard UI. Blue coloured boxes are user inputs, green coloured boxes are LLM outputs. Fully coloured boxes are active as part of the context window, while boxes only with a border represent deactivated context.}
    \label{fig:whitboard-concept}
\end{figure}

The core innovation of our approach is the introduction of a whiteboard-based UI with a context windows (see Figure \ref{fig:whitboard-concept}), which departs from the linear chat paradigm. 
In this system, all user prompts and LLM outputs are represented as discrete, movable elements on a virtual whiteboard. 
This gives the users more granular control over LLM context via the context window (illustrated on the right side of the graphic), where users explicitly control which information is currently active and available to the LLM.

\newpage
We envision the following features for our prototype:

\begin{itemize}
    \item \textbf{Selective Context Management:} Users can add or remove any item (such as prompts, AI responses, or code snippets) to the context window. Only items present in this window are included as context for the next LLM response, enabling more specific interactions.
    
    \item \textbf{Dynamic Reordering:} Items within the context window can be freely rearranged via drag-and-drop (see the stack of blocks in the context window), allowing users to optimize the logical sequence of the provided information.
    
    \item \textbf{Visibility and Control:} The context window remains explicitly visible and manipulable at all times, providing transparency about what the LLM "sees" and ensuring users can continuously adapt the context to their needs.
    
    \item \textbf{Blueprint Design:} Users can save particular arrangements of prompts and responses as reusable templates ("blueprints"), supporting the efficient handling of recurring tasks or workflows.
\end{itemize}

By moving beyond the linear chronological chat interface, our UI empowers users with a higher level of flexibility, transparency, and control in their interactions with LLMs. 
Importantly, the UI is designed for universal compatibility: rather than building a new language model, our focus is on providing a UI that can connect to a wide variety of commercial or open-source LLMs via their APIs. 
This ensures versatility and future-proofing, as users will be able to leverage different underlying AI models depending on their needs.

At present, this  project is in the conceptualization phase where we are systematically exploring and refining the concept. 
Following this phase, we will proceed to prototype development and empirical testing, following the HCD process (see Chapter \ref{ch:hcd}) to ensure the resulting UI is both usable and effective for a diverse range of users.

\subsection{UI Inspirations from the Design Thinking Workshop}
One of the aims of the workshop was to generate further ideas that would be beneficial to our existing UI concept. 
While we analyzed the UI concepts described in Chapter \ref{ch:results}, we found the following ideas particularly valuable for the future development of our UI concept.\\

\noindent
\textbf{Visualization of branching}.
A first area of inspiration concerns the visualization of branching conversation paths. 
The first idea is a path diagram (see Figure \ref{fig:sol_04}), which visually represents the current position within a branched conversation. 
Users can see which path they are on and can easily switch to another branch by clicking within the visualization (similar to minimaps used in computer games). The second idea, inspired by Figures \ref{fig:sol_01} and \ref{fig:sol_06}, is to display conversation branches as horizontally aligned panels on an infinite canvas. 
Every time a new branch is created, a new panel is added to the right, allowing users to scroll left and right to navigate between parallel branches.\\

\noindent
\textbf{Weighting function}.
Another idea is the implementation of a  weighting function (see Figures \ref{fig:sol_04} and \ref{fig:sol_05}) accessible for the user in the UI. 
This could be realized either by highlighting specific text segments within messages or by assigning weights to particular prompts using a scale. 
The weighted inputs would then influence the relative importance of those elements in subsequent LLM outputs, giving users more precise control over how their instructions and feedback shape the AI’s responses.\\

\noindent
\textbf{Hover-based interaction}.
We also see potential in ideas around hover-based interactions for in-message highlights (see Figures \ref{fig:sol_04} and \ref{fig:sol_07}). 
When users highlight parts of a prompt or response, the interface could present alternative formulations (useful for writing support) or offer explanations and tooltips with additional context or clarifications.\\

It is important to mention that it is not necessarily purposeful to integrate all generated ideas into a common prototype: 
While individual ideas may increase the usability of the LLM interface, this does not always mean that in combination they will lead to a good user experience. 
Therefore, our next steps will involve evaluating which concepts best align with our overall design goals and user needs. 
Following the principles of the HCD process, it is useful to develop and test multiple prototypes -- each with a specific set of integrated innovations -- in parallel and test these with users to help identify which ideas deliver the most value in practice. 
This iterative, user-driven selection and refinement will ensure that the resulting UI concept is both innovative and effective.
\section{Discussion}
In this paper, we presented the results of a design thinking workshop held at the deRSE25 conference as part of our development process for innovative UIs for Large Language Models (LLMs). 
We explained the basic functioning of LLMs, highlighted their operational principles, and discussed key limitations of current UIs. 
We then detailed the structure and outcomes of our design thinking workshop, which aimed to collaboratively develop new UI concepts. 
Through participant feedback, we identified common use cases, strengths, and notable weaknesses of LLMs. 
The workshop results, visualized and documented by participants, revealed new UI concepts that offer potential enhancements over current linear chat interfaces, particularly in flexible context management, dynamic branching, and enhanced user control mechanisms. 
These insights now inform our ongoing UI development, reinforcing the value of HCD in optimizing user interaction with AI systems.

The insights derived from our workshop underline that LLMs are extensively utilized both in programming contexts and in everyday tasks. However, due to the nature of our conference audience, there was a distinct emphasis on applications of LLMs in software development (see Chapter \ref{ch:results}). This focus provided rich insight into how users integrate AI tools within programming workflows, with productivity and efficiency gains cited as particularly valuable.

Participants valued the time-saving potential and efficiency offered by LLMs, especially their ability to handle repetitive and monotonous tasks, such as writing documentation for coding projects. However, significant concerns were also expressed, particularly regarding the reliability of outputs. Users frequently mentioned issues such as inaccuracies and occasional nonsensical outputs, which undermined their trust and required careful verification.

Interestingly, while these criticisms were abundant, specific critiques of the UI were notably scarce. Apart from a solitary mention regarding the inconvenience of manually copying content between chat instances, participants generally accepted existing UI conventions without substantial complaint.

This lack of explicit UI criticism suggests that existing interfaces are broadly functional, but not necessarily optimal. We propose that the relative absence of UI-focused criticism does not indicate complete satisfaction but rather reflects users' habituation to current interaction paradigms. We also deliberately asked for all kinds of criticism -- not only concerning the UI -- in order to avoid restricting participants' feedback to a narrow topic too early in the development process.

Beyond these points, participants also raised important ethical and practical considerations, including data security, privacy, and the social and environmental impacts associated with AI. Concerns were particularly voiced regarding data privacy risks, potential misuse of data by companies, and broader social consequences, such as labor exploitation in the training of AI models. Environmental concerns were also prominent, highlighting the significant resource consumption and ecological footprint associated with training large-scale AI models. While these concerns cannot be addressed solely through alternative UIs, they highlight the need for broader social discourse about the implications of AI and the necessity for interdisciplinary collaboration to address these challenges in the future.

For us, the HCD process has proven to be a particularly valuable framework for structuring our development efforts, even though our work is still in the conceptual phase. Its iterative nature and explicit focus on user involvement enable a robust alignment of our UI concepts with user needs and expectations. Moving forward, we will refine our initial UI concepts by integrating the ideas generated during the workshop. Subsequently, we will create wireframe prototypes and engage in multiple iterations guided by user evaluations. Through this development process, we aim to arrive at a fully functional UI, which we intend to share with the broader scientific community.

Underlying our future development effort is the following belief: We hypothesize that, although LLM capabilities have seen rapid advancement, characterized by increased parameters and expanded context windows, technical progress is likely to plateau in the near future. At this stage, interface usability and user experience will become decisive differentiators between competing models. We therefore emphasize that future development of LLMs should increasingly prioritize UI innovation, ensuring that UI design receives the attention necessary to fully exploit the potential of AI systems.

\begin{acknowledge}
\\
\\
\textbf{Funding} \\
We thankfully acknowledge financial support by the Münchner Universitätsgesellschaft awarded as part of the AI-HUB@LMU Prize for the Most Innovative AI-based Research Project.\\

\noindent
\textbf{Competing Interests} \\
The authors declare no competing interests. \\

\noindent
\textbf{Author Contributions} \\
All author meet the criteria for authorship. The specific contributions are described below using the CRediT taxonomy (\url{https://credit.niso.org}).

\vspace{0.3cm}
\begin{tabular}{ll}
Conceptualization: & Maximilian Frank, Simon Lund \\
Methodology: & Maximilian Frank, Simon Lund \\
Visualization: & Maximilian Frank, Simon Lund \\
Writing – original draft: & Maximilian Frank, Simon Lund\\
Funding acquisition: & Maximilian Frank, Simon Lund \\ 
\end{tabular}
\end{acknowledge}

\bibliographystyle{eceasst}
\bibliography{deRSE-workshop-dynamic-llm-interfaces/eceasst}

\end{document}